\documentclass[9pt,conference,a4paper]{IEEEtran}
\ifCLASSINFOpdf
\else
\fi
\usepackage{amssymb}
\usepackage{bm}
\usepackage[french,english]{babel}
\usepackage{color}
\usepackage{amsmath}
\usepackage[]{graphicx}

\begin{document}

\title{Why CLEAN when you can PURIFY? A new approach for next-generation radio-interferometric imaging}

\author{%
\IEEEauthorblockN{
Rafael E. Carrillo\IEEEauthorrefmark{1}, 
Jason D. McEwen\IEEEauthorrefmark{2}, 
and Yves Wiaux\IEEEauthorrefmark{3}}
\IEEEauthorblockA{\IEEEauthorrefmark{1} 
Institute of Electrical Engineering, Ecole Polytechnique F{\'e}d{\'e}rale de Lausanne (EPFL),
      CH-1015 Lausanne, Switzerland.}
\IEEEauthorblockA{\IEEEauthorrefmark{2}
Mullard Space Science Laboratory, University College London (UCL), Holmbury St Mary, Surrey RH5 6NT, UK.}
\IEEEauthorblockA{\IEEEauthorrefmark{3}
Institute of Sensors, Signals, and Systems, Heriot-Watt University, Edinburgh EH14 4AS, UK.}
}

\maketitle

\begin{abstract}
In recent works, sparse models and convex optimization techniques have been applied to radio-interferometric (RI) imaging showing the potential to outperform state-of-the-art imaging algorithms in the field. In this talk, I will review our latest contributions in RI imaging, which leverage the versatility of convex optimization to both handle realistic continuous visibilities and offer a highly parallelizable structure paving the way to high-dimensional data scalability. Firstly, I will review our recently proposed average sparsity approach, SARA, which relies on the observation that natural images exhibit strong average sparsity over multiple coherent bases. Secondly, I will discuss efficient implementations of SARA, and sparse regularization problems in general, for large-scale imaging problems in a new toolbox dubbed PURIFY. 
\end{abstract}
\vspace*{9pt}
The advent of next-generation radio telescopes, such as the new LOw Frequency ARray (LOFAR), the recently upgraded Karl G. Jansky Very Large Array (VLA) and the future Square Kilometer Array (SKA), has posed several challenges for image reconstruction and the design of data processing systems \cite{wijnholds14}. The new telescopes will achieve much higher dynamic range than current instruments at a higher angular resolution. Also, these telescopes will acquire a massive amount of data, thus posing large-scale inverse problems in the perspective of image reconstruction. These challenges have triggered an intense research in the community to reformulate imaging and calibration techniques for radio interferometry (RI).

The RI measurement equation can be discretized as $\bm{y}=\mathsf{\Phi}\bm{x}+\bm{n}$, where $\bm{y}\in\mathbb{C}^{M}$ denotes the vector of measured visibilities, $\mathsf{\Phi}\in\mathbb{C}^{M\times N}$ is a discretization of the measurement operator and $\bm{n}\in\mathbb{C}^{M}$ represents the observation noise. In \cite{carrillo12} we propose an imaging algorithm dubbed sparsity averaging reweighted analysis (SARA) based on average sparsity over multiple bases, showing superior reconstruction qualities relative to state-of-the-art imaging methods in the field. A sparsity dictionary composed of a concatenation of $q$ coherent bases, 
$\mathsf{\Psi}=[\mathsf{\Psi}_1, \mathsf{\Psi}_2, \ldots, \mathsf{\Psi}_q]$,
is used and average sparsity is promoted through the minimization of an analysis $\ell_0$ prior, $\|\mathsf{\Psi}^{\dagger}\bar{\bm{x}}\|_{0} $, where $\mathsf{\Psi}^{\dagger}$ denotes the adjoint operator of $\mathsf{\Psi}$ \cite{carrillo13}. 

SARA adopts a reweighted $\ell_1$ minimization scheme to promote average sparsity through the prior $\|\mathsf{\Psi}^{\dagger}\bar{\bm{x}}\|_{0} $. The algorithm replaces the $\ell_0$ norm by a weighted $\ell_1$ norm and solves a sequence of weighted $\ell_1$ problems where the weights are essentially the inverse of the values of the solution of the previous problem \cite{carrillo12,carrillo13}. The weighted $\ell_1$ problem is defined as:
\begin{equation}\label{delta}
\min_{\bar{\bm{x}}\in\mathbb{R}_{+}^{N}}\|\mathsf{W\Psi}^{\dagger}\bar{\bm{x}}\|_{1}
\textnormal{ subject to }\| \bm{y}-\mathsf{\Phi}\bar{\bm{x}}\|_{2}\leq\epsilon,
\end{equation}
where $\mathsf{W}\in\mathbb{R}^{D\times D}$ denotes the diagonal matrix with positive weights, $\mathbb{R}^{N}_{+}$ denotes the positive orthant in $\mathbb{R}^{N}$ and $\epsilon$ is an upper bound on the $\ell_{2}$ norm of the noise, which can be accurately estimated. Hence, we focus our attention on solving problem \eqref{delta} efficiently, especially for large-scale data problems, i.e. when the number of visibilities is very large ($M\gg N$). In this case, we propose to split the data vector $\bm{y}$ and the measurement operator into $R$ blocks in the following manner:
\begin{equation}\label{blocks}
\bm{y}= [\bm{y}_1^T,\dots,\bm{y}_R^T]^T \textnormal{ and  } \mathsf{\Phi}= [\mathsf{\Phi}_1^T, \dots, \mathsf{\Phi}_R^T]^T,
\end{equation}
where each $\bm{y}_i$ is modelled as $\bm{y}_i=\mathsf{\Phi}_i \bm{x} + \bm{n}_i$ and $\bm{n}_i$ denotes the noise vector. With this partition the optimization problem in \eqref{delta} can be reformulated as
\begin{equation}\label{dbp1}
\min_{\bar{\bm{x}}\in\mathbb{R}_{+}^{N}}\|\mathsf{W\Psi}^{\dagger}\bar{\bm{x}}\|_{1}
\textnormal{ subject to }\| \bm{y}_i-\mathsf{\Phi}_i\bar{\bm{x}}\|_{2}\leq\epsilon_i, ~i=1,\dots,R,
\end{equation}
where each $\epsilon_i$ is an appropriate bound for the $\ell_2$ norm of $\bm{n}_i$.

In \cite{carrillo14} we propose a general algorithmic framework based on the simultaneous-direction method of multipliers (SDMM) to solve \eqref{dbp1}. The proposed framework offers a parallel implementation structure that decomposes the original problem into several small simple problems, hence allowing implementation in multicore architectures or in computer clusters, or on graphics processing units. These implementations provide both flexibility in memory requirements and a significant gain in terms of speed, thus enabling scalability to large-scale problems. A beta version of an SDMM-based imaging software written in C and dubbed PURIFY was released that handles various sparsity priors, including SARA, thus providing a new powerful framework for RI imaging (toolbox available at http://basp-group.github.io/purify/). Even though this beta version of PURIFY is not parallelized yet, we discuss in detail the extraordinary parallel and distributed optimization potential of SDMM to be exploited in future versions. We also discuss other possible research avenues for big data scalability. One possibility is to incorporate ideas from stochastic gradient methods into proximal splitting and augmented Lagrangian methods. The key idea is to use only one data block $\bm{y}_i$, or a subset of blocks, at each iteration of the reconstruction. By doing so, the computational complexity per iteration will be reduced. Thus, the total processing time of the algorithm will also be reduced if the convergence rate of the original problem is preserved. See for example \cite{azadi14} and references therein for first theoretical results. 

\bibliographystyle{IEEEtran}
\bibliography{abrev,sara}

\end{document}